\documentclass[nofootinbib]{revtex4}
\input psfig.sty
\usepackage[dvips]{color}
\usepackage{graphicx}
\usepackage{epsfig}

\begin{document}

\title{A cosmological concordance model with dynamical vacuum term}

\author{J. S. Alcaniz$^1$\email{alcaniz@on.br}, H. A. Borges$^{2}$\email{humberto@ufba.br}, S. Carneiro$^{2}$\footnote{ICTP Associate Member.}\email{saulo.carneiro@pq.cnpq.br}, J. C. Fabris$^3$\email{fabris@pq.cnpq.br}, C. Pigozzo$^{2}$\email{kssiobr@gmail.com}, W. Zimdahl$^{3}$\email{zimdahl@online.de}}

\affiliation{$^1$Observat\'orio Nacional, Rio de Janeiro, RJ, Brazil\\$^2$Instituto de F\'{\i}sica, Universidade Federal da Bahia, Salvador, BA, Brazil\\$^3$Departamento de F\'{\i}sica, Universidade Federal do Esp\'{\i}rito Santo, Vit\'oria, ES, Brazil}

\date{\today}

\begin{abstract}
We demonstrate that creation of dark-matter particles at a constant rate implies the existence of a cosmological term that decays linearly with the Hubble rate.
We discuss the cosmological model that arises in this context and test it against observations of the  first acoustic peak in the cosmic microwave background (CMB) anisotropy spectrum, the Hubble diagram for supernovas of type Ia (SNIa), the distance scale of baryonic acoustic oscillations (BAO) and the distribution of large scale structures (LSS). We show that a good concordance is obtained, albeit with a higher value of the present matter abundance than in the $\Lambda$CDM model. We also comment on general features of the CMB anisotropy spectrum and on the cosmic coincidence problem.
\end{abstract}

\maketitle

\section{Introduction}

The cosmological constant problem is usually described as a huge difference between the vacuum energy density derived by quantum field theories and the observed value of the cosmological constant. The energy density associated to the vacuum fluctuations of free fields is a divergent quantity, and any natural cutoff we use to regularize it (for example, the Planck mass or the energy scale of the quantum chromodynamic (QCD) phase transition, the latest cosmological vacuum transition) leads to values many orders of magnitude above the observed $\Lambda$. But the applicability of flat-spacetime quantum field theory in the expanding universe is doubtful, since a non-vanishing bare cosmological constant is not compatible with a flat spacetime.
The vacuum contribution to (semi-classical) gravity is only properly calculated when one considers a non-flat background and renormalizes the divergent contributions in some appropriate way. This is a difficult task and, in general, may depend on the renormalization method used. In the simpler case of free conformal fields in a de Sitter spacetime \cite{H4}, the renormalized vacuum density is proportional to $H^4$, where $H$ is the Hubble parameter. In the high-energy limit this result can be used to obtain a non-singular cosmology, with an initial quasi-de Sitter phase, giving origin to a subsequent radiation dominate period \cite{Tavakol,Carneiro}. However, in the present universe the above result leads to a very tiny vacuum term. On the other hand, it may well be possible that the observed cosmological term is just a cosmological constant of purely geometric nature. In this case there would be no natural (micro-) physical scale associated to it. Furthermore, both the gravitational action of the quantum vacuum and the approximate coincidence between such a constant and the present matter density (the cosmic coincidence problem) would remain unexplained. 

As long as this issue cannot be solved on a fundamental level, (semi-) phenomenological models which avoid the theoretical difficulties pointed out above may help to get further insight into the problem.  In this letter we argue that, by modeling the cosmological “constant” in terms of a decaying vacuum energy, it is possible to obtain a model which is competitive with the standard one from the observational point of view.  
A crucial point here is to consider the contribution of interacting fields to the vacuum energy. Indeed, some authors have been arguing that the energy density associated to the QCD vacuum condensate in a low-energy de Sitter (or approximately de Sitter) spacetime is given by
\begin{equation} \label{Schutz}
\Lambda \approx m^3 H,
\end{equation}
where $m \approx 150$ MeV is the energy scale of the QCD phase transition \cite{QCD}. In the present universe this density is dominant when compared to the contribution of free fields, $H^4$, and has the correct order of magnitude of the observed cosmological term. 
In this paper we will trace back this scaling law $\Lambda\propto H$ to a process of matter production in the expanding spacetime.

\section{Particle production and the vacuum density}

Let us consider a low-energy, spatially flat Friedmann-Lema\^{\i}tre-Robertson-Walker (FLRW) spacetime, where non-relativistic dark particles of mass $M$ are created out of the vacuum at a constant rate $\Gamma$. The particle number balance (sometimes called Boltzman equation) for this process will be given by
\begin{equation} \label{Boltzman}
\frac{1}{a^3}\frac{d}{dt}\left(a^3n\right)= \Gamma n,
\end{equation}
where $n$ is the particle number density at a given time. It can be rewritten as
\begin{equation} \label{conservacao}
\dot{\rho}_m + 3H\rho_m = \Gamma \rho_m,
\end{equation}
where $\rho_m = nM$  is the matter density. This equation is characteristic of models with interaction in the dark sector \cite{winfried,Chimento,delCampo,Sola}.

This particle production from vacuum fluctuations necessarily involves a gravitational backreaction. In addition to equation (\ref{conservacao}) we have the Friedmann equation
\begin{equation} \label{Friedmann}
\rho = \rho_m + \Lambda = 3H^2,
\end{equation}
where $\Lambda$ is the vacuum contribution to the energy density. We shall assume $p_{\Lambda} = - \Lambda$ for the vacuum pressure, admitting, however, $\Lambda$ to be time varying. The total energy density satisfies the conservation equation
\begin{equation} \label{conservacao2}
\dot{\rho} + 3H(\rho +p) = 0,
\end{equation}
provided we take $\Lambda = 2\Gamma H + \lambda_0$, where $\lambda_0$ is a constant of integration, which corresponds to a purely geometric cosmological constant. Since there is no natural (microphysical) energy scale associated to this constant, it is disregarded here. In this way we have
\begin{equation} \label{linear}
\Lambda = 2\Gamma H,
\end{equation}
and equation (\ref{conservacao2}) can be rewritten as
\begin{equation}
\dot{\rho}_m + 3H\rho_m = -\dot{\Lambda}.
\end{equation}

We can see that the particle creation is concomitant with a decay in the vacuum energy density. Equation (\ref{linear}) is the same time variation law (\ref{Schutz}) obtained in a quasi-de Sitter spacetime when the contribution of the QCD condensate is taken into account \cite{QCD}. In this case we have $\Gamma \approx m^3$, where $m$ is the energy scale of the QCD phase transition\footnote{We are using natural units with $8 \pi G = c = \hbar = 1$.}. As discussed in \cite{QCD}, the association of the QCD energy scale with the observed vacuum energy is natural and, in some way, expected, since the QCD confinement, with breaking of chiral symmetry, is the last cosmological vacuum transition we know. The QCD condensate formed after this low energy transition gives, in a de Sitter (or approximately de Sitter) spacetime, the correct value of the cosmological term (see below). Furthermore, owing to the maximal symmetry of the de Sitter background, the condensate pressure is the negative of its energy density, i.e., the same equation of state associated to a cosmological constant.

Dividing equation (\ref{linear}) by $3H^2$, using $\Lambda/(3H^2) \equiv 1 - \Omega_m$ and specifying to the present time (subscripts $0$), we obtain
\begin{equation} \label{rate}
\Gamma = \frac{3}{2} \left( 1 - \Omega_{m0} \right) H_{0}.
\end{equation}
In a de Sitter universe ($\Omega_m = 0$) we would have $\Gamma = 3H/2$. That is, the particle creation rate would be equal (apart from a numerical factor) to the Gibbons-Hawking temperature associated to the event horizon \cite{Gibbons}. On the other hand, by taking $\Omega_{m0} \approx 1/3$ for our present universe, we have $H_0 \approx m^3$ and $\Lambda \approx m^6$. The former relation is an expression of the old Eddington-Dirac large number coincidence \cite{Mena}, while the latter (known as Zeldovich's relation \cite{Bjorken}) gives the correct order of magnitude for $\Lambda$. The exact value of $\Gamma$ in (\ref{rate}) (and hence of $\Lambda_0$) will be given by observations in the next sections. Once $\Gamma$ is fixed, no further fine-tuning will be necessary. The cosmological model
to be built will have the same number of free parameters as  the  $\Lambda$CDM model, namely $\Omega_{m0}$ and $H_0$.

\section{The model}

From this section on we will analyze the cosmological model arising from a vacuum energy density scaling with time in accordance with the ansatz discussed above. When compared with the most precise observational constraints, the model will show a very good concordance and, in some cases, less tensions than in the standard case. Therefore, besides providing a possible physical basis for the cosmological term, in the lines of the previous sections, it will also show to be competitive from the observational point of view.

With $\Lambda = 2\Gamma H$ we obtain, from the Friedmann equations, the solution \cite{Humberto,Jailson}
\begin{equation}\label{Hgeral}
\frac{H}{H_0} \approx \left\{ \left[1 - \Omega_{m0} + \Omega_{m0} (1
+ z)^{3/2}\right]^2 + \Omega_{r0} (1+z)^4 \right\}^{1/2}\;,
\end{equation}
where $\Omega_{m0}$ is the present relative matter density, and we have added conserved radiation with present density parameter $\Omega_{r0}$. As discussed in \cite{Jailson,Jailson2,Jailson3}, for non-zero $\Omega_{r0}$ the expression (\ref{Hgeral}) is an approximate solution, differing only $1\%$ from the exact one, since $\Omega_{r0} \approx 8 \times 10^{-5} \ll 1$. For $\Omega_{r0} = 0$, the solution (\ref{Hgeral}) is exact.

For early times we obtain $H^2(z) = H_0^2 \Omega_{r0} z^4$, and the radiation era is indistinguishable from the standard one. On the other hand, for high redshifts the matter density scales as $\rho_m(z) = 3H_0^2 \Omega_{m0}^2 z^3$. The extra factor $\Omega_{m0}$ - as compared to the $\Lambda$CDM model - is owing to the late-time process of matter production. In order to have nowadays the same amount of matter, we need less matter in the past. Or, in other words, if we have the same amount of matter in the past (say, at the time of matter-radiation equality), this will lead to more matter today. We can also see from (\ref{Hgeral}) that, in the asymptotic limit $z \rightarrow -1$, the solution tends to the de Sitter solution.

Note that, like the $\Lambda$CDM model, the above model has only two free parameters, namely $\Omega_{m0}$ and $H_0$. On the other hand, it can not be reduced to the $\Lambda$CDM case
except for $z\rightarrow -1$. In this sense, it is falsifiable, that is, it may be ruled out by observations.

\section{Background tests}

The Hubble function (\ref{Hgeral}) is all we need to test the model against observations of SNIa and BAO \cite{Jailson,Jailson2,Jailson3}. However, when testing the position of the first peak of the CMB anisotropy spectrum, some care is needed. The relation between the observed position of the peak and the acoustic scale $l_A$ is given by \cite{Tegmark}
\begin{equation}\label{l1}
l_1 = l_A (1 - \delta_1), \quad  \mbox{where} \quad \delta_1 = 0.267 \left( \frac{r}{0.3} \right)^{0.1},
\end{equation}
with $r \equiv \rho_r/\rho_m$ evaluated at the redshift of last scattering $z_{ls}$. The acoustic scale is defined, as usual, by
\begin{equation}\label{lA}
l_A = {\pi \int_0^{z_{ls}}
\frac{dz}{H(z)}}/{\int_{z_{ls}}^{\infty}\frac{c_s}{c}\frac{dz}{H(z)}},
\end{equation}
with the sound velocity
\begin{equation}\label{cs}
c_s = c \left( 3 + \frac{9}{4}\frac{\Omega_{b0}}{z\Omega_{\gamma 0}}
\right)^{-1/2},
\end{equation}
where $\Omega_{b0}$ and $\Omega_{\gamma 0}$ stand for the present density parameters of baryons and photons, respectively. The relation (\ref{l1}) does not depend on the dark energy model, but only on pre-recombination physics. However, the ratio $r$ is, in our case, given by
\begin{equation} \label{r}
r = \frac{\Omega_{r0}}{\Omega_{m0}^2} z_{ls},
\end{equation}
with the extra factor $\Omega_{m0}$ as compared to the $\Lambda$CDM expression. Note that, in all the above expressions, there is a dependence on $H_0$ through the relative densities of baryons, photons and radiation, as well as through the redshift of last scattering. In our analysis we will fix these parameters ($c_{s}$, $\Omega_{r0}$, $z_{ls}$) to the concordance values of the standard model. The only free parameters to be adjusted will then be $\Omega_{m0}$ and $H_0$.

\begin{figure*}
\vspace{.2in}
\centerline{\psfig{figure=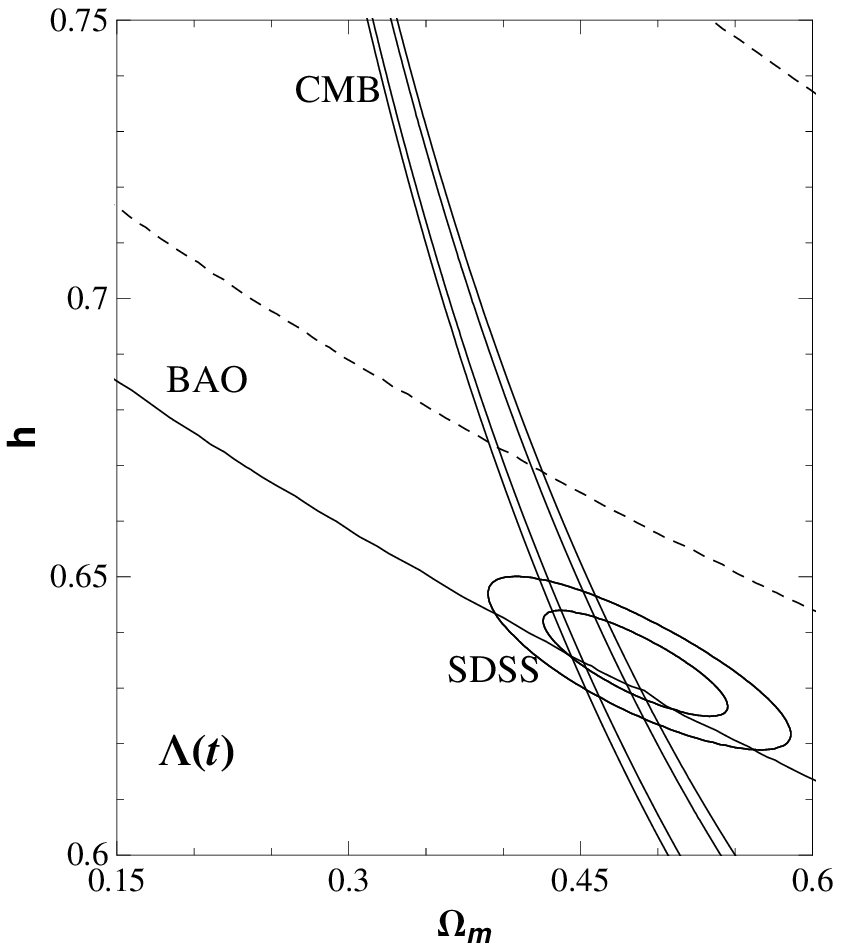,width=2.2truein,height=2.2truein}
\psfig{figure=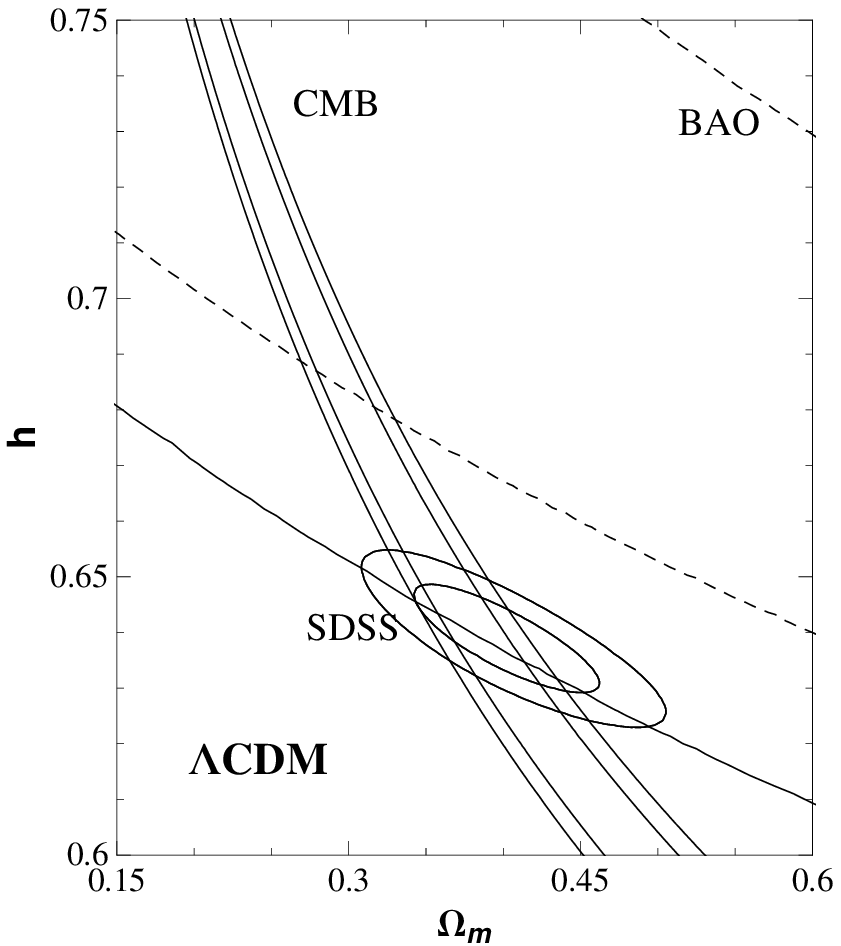,width=2.2truein,height=2.2truein}
\psfig{figure=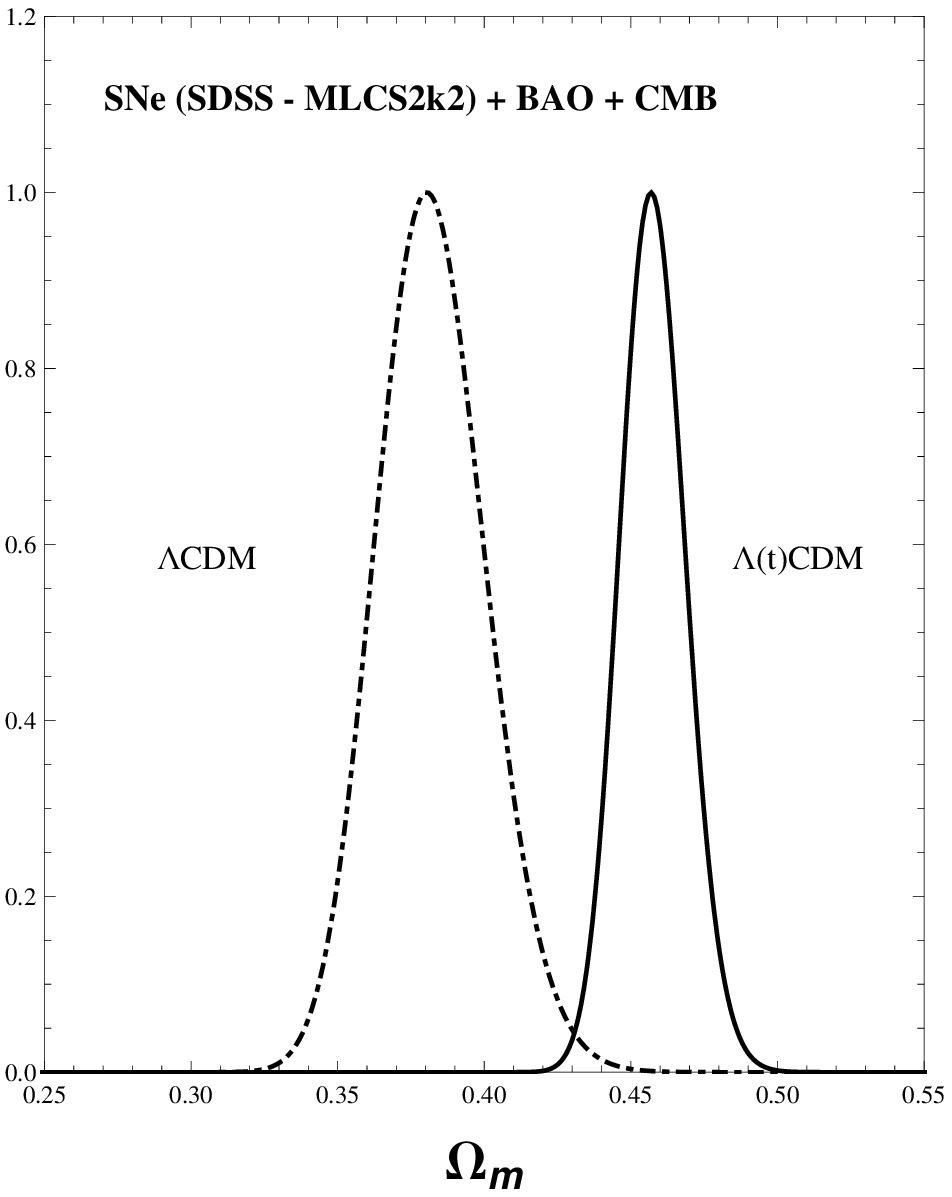,width=2.2truein,height=2.2truein}}
\caption{{\bf Left panel:} superposition of the confidence regions from SDSS (MLCS2k2) SNIa, BAO and the position of the first peak of the CMB in the present model \cite{ Jailson3}. {\bf Center panel:} the same for the spatially flat $\Lambda$CDM. {\bf Right panel:} The corresponding likelihoods when $h$ is marginalized.}
\label{figsdss}
\end{figure*}

The result of a joint analysis of SNIa, BAO and the position of the first peak of the CMB spectrum is shown in Figure \ref{figsdss}. In the left panel we show the superposition of the corresponding confidence regions in the ($h$,$\Omega_{m0}$) plane, where $h = H_0/100$(km/s)/Mpc is the dimensionless  Hubble constant. The central panel shows the same for the $\Lambda$CDM model. In the right panel a comparison between the two models is made, after marginalizing over $h$. In this analysis we have used the SDSS SNIa compilation with the MLCS2k2 light-curve fitter \cite{SDSS}. Other current compilations make use of Salt or Salt2 fitters which, however, are not model independent. A more complete analysis, including the Union2 and Constitution compilations, can be found in \cite{Jailson3}. In the case of the Constitution sample calibrated with MLCS2k2, the results are very similar to those presented here. Note that for the standard model the relative matter density is higher than the usually accepted concordance value ($\Omega_{m0}\approx 0.27$). This is a characteristic of the MLCS2k2 fitter, also present when we use it to calibrate the Constitution sample.

\begin{figure*}[]
\vspace{.2in}
\centerline{
\psfig{figure=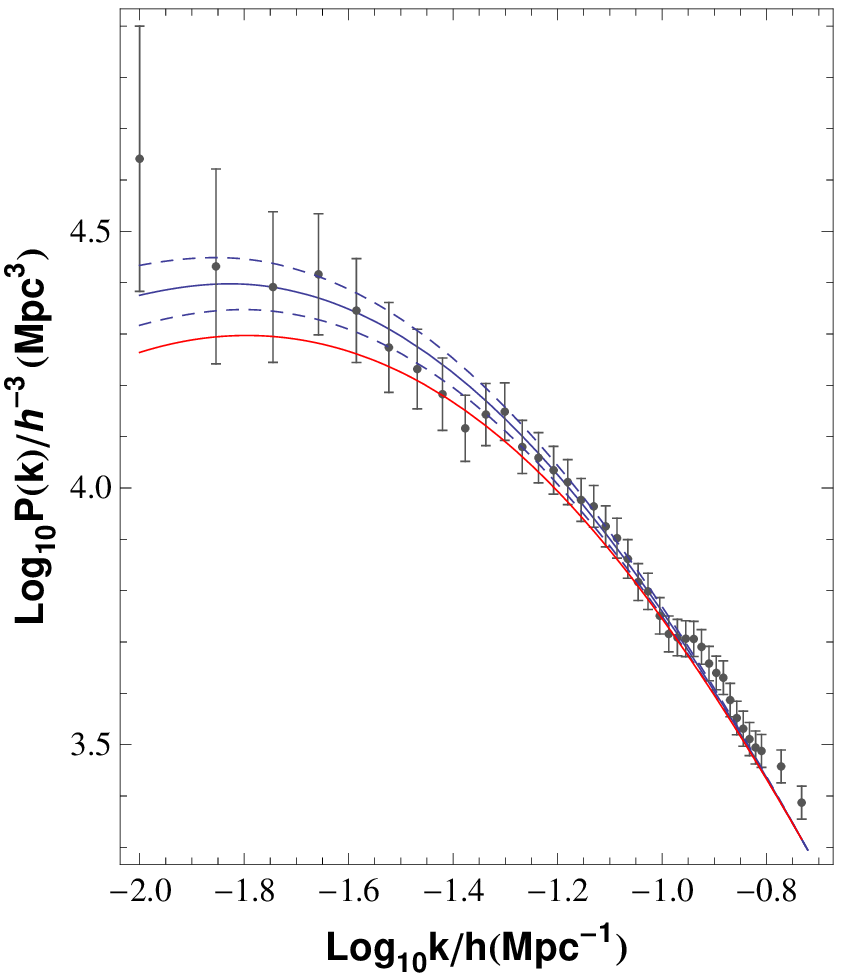,width=2.2truein,height=2.15truein}\hspace{.6in}
\psfig{figure=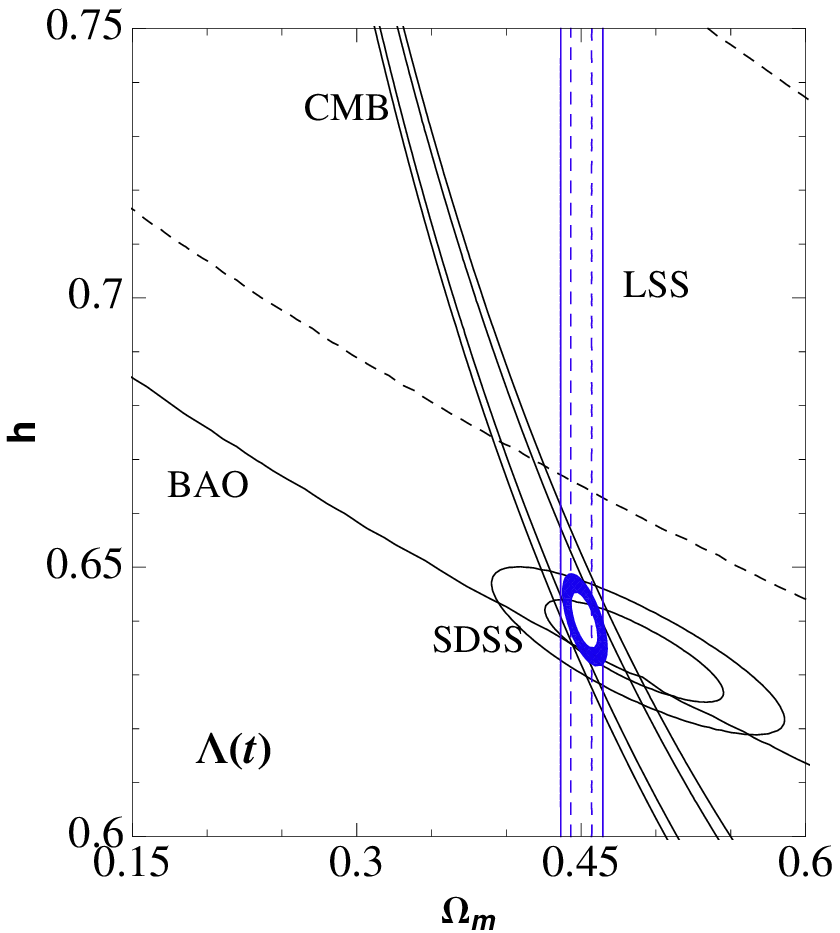,width=2.2truein,height=2.2truein}}
\caption{{\bf Left panel:} the 2dFGRS power spectrum in the present model with $\Omega_{m0} = 0.45$ (blue solid line) and for the BBKS transfer function with $\Omega_{m0} = 0.27$ (red line). {\bf Right panel:} superposition of the four tests (SNIa, BAO, CMB and LSS) performed with the present model. The blue ellipse corresponds to the concordance region.}
\label{figpowersp}
\end{figure*}

\section{The matter power spectrum}

The analysis of the matter power spectrum was performed in Ref. \cite{Julio}, where, for simplicity, baryons were not included and the cosmological term was not perturbed. In a subsequent publication a gauge-invariant analysis, explicitly considering the presence of late-time non-adiabatic perturbations, has shown that the vacuum perturbations are indeed negligible, except for scales near the horizon \cite{Zimdahl}. Therefore, the result of Ref. \cite{Julio} can be considered a good approximation. An updated analysis is shown in the left panel of Figure \ref{figpowersp}, together with the 2dFGRS data points \cite{2dF} and, for comparison, the concordance $\Lambda$CDM fitting as provided by the BBKS transfer function \cite{BBKS}. It is also shown the $2\sigma$ confidence region for our fitting.

\section{Joint analysis}

\begin{table}[t]
\begin{center}
\caption{Limits to $\Omega_{m0}$ (SNe+ CMB + BAO+LSS).}
\begin{tabular}{rcccc}
\hline \hline \\
\multicolumn{1}{c}{ } & \multicolumn{2}{c}{$\Lambda(t)$CDM } & \multicolumn{2}{c}{$\Lambda$CDM } \\
\multicolumn{1}{c}{Test}&
\multicolumn{1}{c}{$\Omega_{m0}$\footnote{Error bars stand for $2\sigma$}}&
\multicolumn{1}{c}{$\chi^2_{min}/\nu$}&
\multicolumn{1}{c}{$\Omega_{m0}$$^a$}&
\multicolumn{1}{c}{$\chi^2_{min}/\nu$}\\ \hline \\
Union2 (SALT2)......&$0.420^{+0.009}_{-0.010}$ & 1.063 & $0.235\pm 0.011$ & 1.027 \\
SDSS (MLCS2k2).......& $0.450^{+0.014}_{-0.010}$ & 0.842 & $0.260^{+0.013}_{-0.016}$ & 1.231 \\
Constitution (MLCS2k2[17]).......& $0.450^{+0.008}_{-0.014} $ &1.057 & $0.270\pm 0.013$ & 1.384\\
\hline \hline
\end{tabular}
\end{center}
\end{table}

\begin{figure*}[]
\vspace{.2in}
\centerline{
\psfig{figure=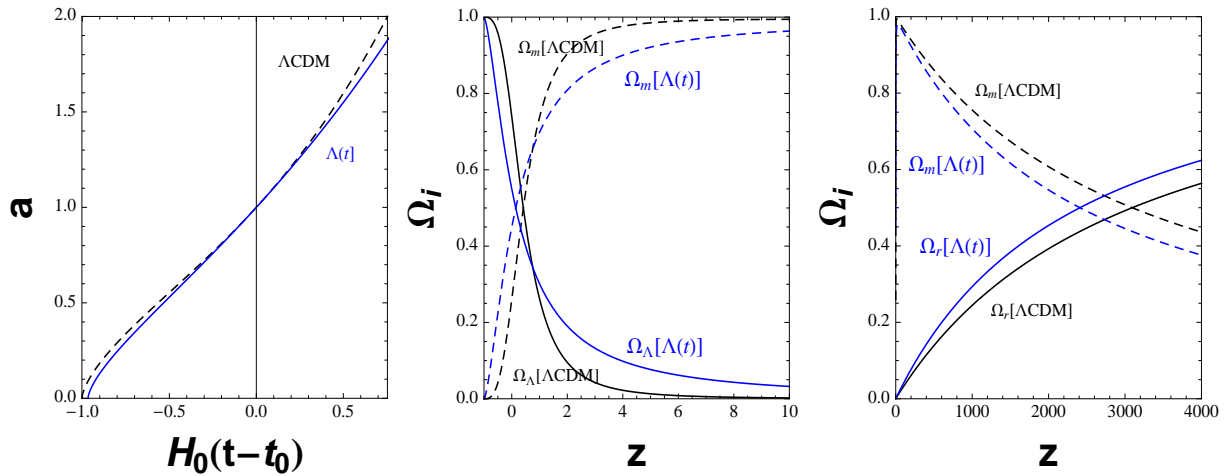,width=6.4truein,height=2.5truein}}
\caption{{\bf Left panel:} The time evolution of the scale factor in the spatially flat $\Lambda$CDM and in the present model. {\bf Center panel:} The density parameters of matter and dark energy in both models. {\bf Right panel:} The density parameters of matter and radiation in both models. In all cases we are using the best-fit values of $\Omega_{m0}$ from the SDSS analysis, and the density parameters are defined as $\Omega_i = \rho_i(z)/3H^2(z)$.}
\label{figure03}
\end{figure*}

In the right panel of Figure \ref{figpowersp} we show the superposition of the $1\sigma$ and $2\sigma$ confidence regions for the four tests we are considering here (SNe Ia, BAO, first peak of CMB and LSS), as well as the concordance ellipses of our joint analysis. The concordance value for the matter density is $0.44 < \Omega_{m0} < 0.46$ ($2\sigma$). The Hubble parameter is smaller than in other estimations, but it has the same value obtained when the same SNIa sample is analyzed for the $\Lambda$CDM model (see Fig. \ref{figsdss}). In what concerns the matter density parameter, it is about $10\%$ higher than the flat standard model value for the SDSS (MLCS2k2) compilation ($\Omega_{m0} \approx 0.40$). In both models the obtained values of $\Omega_{m0}$ are higher than the usually accepted concordance value for the flat $\Lambda$CDM model, and the same occurs when one uses the Constitution sample calibrated with MLCS2k2. A discussion about the present tension between different SNIa samples and fitters, in both models, can be found in \cite{Jailson3}. Here we show, in Table I, the best-fit results for $\Omega_{m0}$ (with $h$ marginalized) with three samples of supernovas: the SDSS and Constitution calibrated with the MLCS2k2 fitter, and the Union2 (which is calibrated with SALT2). In Figure \ref{figure03}, on the other hand, the reader can find the evolution of the scale factor (left panel) and of the relative energy densities of radiation, matter and dark energy (central and right panels) in the two models.

The concordance value obtained for $\Omega_{m0}$ is in agreement with estimates based on peculiar velocity measurements for galaxy pairs \cite{galaxies}, but only in marginal agreement with dynamical estimations from clusters \cite{clusters}. It is important to emphasize, however, that such estimations are not very precise and that they involve a different scale.
Note also that the present result has an approximative character. As discussed in \cite{Julio}, the inclusion of baryons in the analysis of the power spectrum may lead to a difference of about $10\%$ in the estimation of $\Omega_{m0}$.

With the concordance values of $h$ and $\Omega_{m0}$ in hand, we can obtain the age of the Universe, as well as the redshift of transition between the decelerated and accelerated phases. They are given, respectively, by \cite{Jailson}
\begin{equation} \label{idade}
H_0t_0 = \frac{2\ln\Omega_{m0}}{3(\Omega_{m0}-1)} ,
\end{equation}
and
\begin{equation} \label{transicao}
z_T = \left[2\left(\frac{1}{\Omega_{m0}}-1\right)\right]^{2/3}-1.
\end{equation}
This leads to $H_0 t_0 = 0.97$ and $z_T = 0.81$, in good agreement with standard predictions and astronomical limits \cite{age}. For $h \approx 0.7$ we have $t_0 \approx 13.5$ Gyr.

\section{The CMB anisotropy spectrum}

It is not easy to perform a full analysis of the CMB anisotropy spectrum with alternative models, specially if it involves new processes like particle production, for example. Nevertheless,
for at least some of the features
we do not expect significant changes.
First of all, radiation and baryons are assumed to be separately conserved (there is no photon fluctuations from the vacuum at the tree level, and baryons are too heavy to be produced at late-times). Therefore, the CMB temperature, the acoustic horizon at the last scattering and the relative height of the peaks will not be affected.

On the other hand, the redshift of equality between matter and radiation in our model is given by
\begin{equation} \label{igualdade}
z_{eq} = \frac{\Omega_{m0}^2}{\Omega_{r0}},
\end{equation}
where we can note again an extra factor $\Omega_{m0}$ as compared to the standard $\Lambda$CDM result. By using our concordance matter density $\Omega_{m0} = 0.45$, we obtain $z_{eq} \approx 2500$, which does not differ too much from the standard value. (This is simply because $0.25 = 0.5^2$. Therefore, a relative matter density of around $0.5$ in the present model leads to the same $z_{eq}$ of the standard model with $\Omega_m \approx 0.25$). For this same reason, the matter density at the time of equality will be approximately the same in both models. [By the way, this result for $z_{eq}$ guarantees that the turnover of the matter power spectrum is correctly placed (see Fig. \ref{figpowersp} and \cite{Julio})]. As the redshift of last scattering is exactly the same \cite{Jailson2}, and the recombination does not depend on a particular dark energy model, both last scattering and recombination will occur in the matter dominated phase. All these results indicate that the absolute height of the peaks will not change either.\footnote{In reference \cite{chineses} the authors claim, by using an adapted version of CAMB, that a good fitting for the CMB spectrum can be obtained with our concordance matter density, $\Omega_{m0} \approx 0.45$ (see the left panel of their Figure 3). On the other hand, they claim that this density leads to a matter power spectrum in disagreement with LSS data, contrary to what we have obtained in \cite{Julio}. Our present results, however, agree with those obtained previously in reference \cite{Julio}.}

\begin{figure*}[]
\vspace{.2in}
\centerline{
\psfig{figure=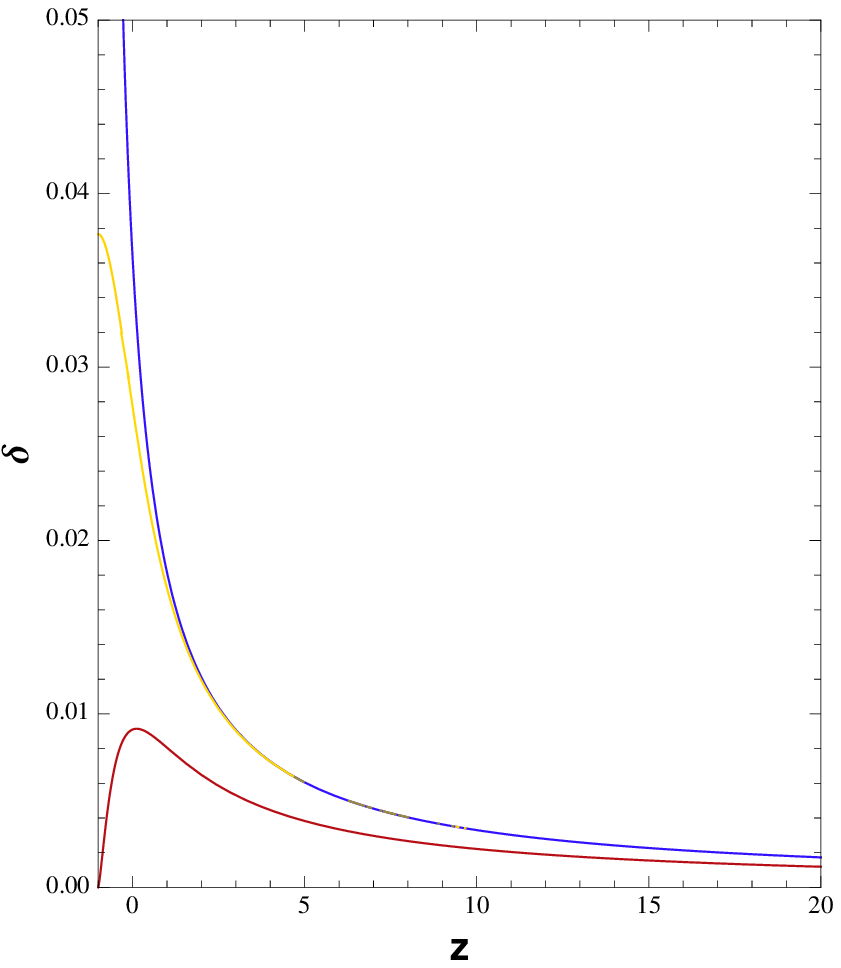,width=2.2truein,height=2.15truein}\hspace{.6in}
\psfig{figure=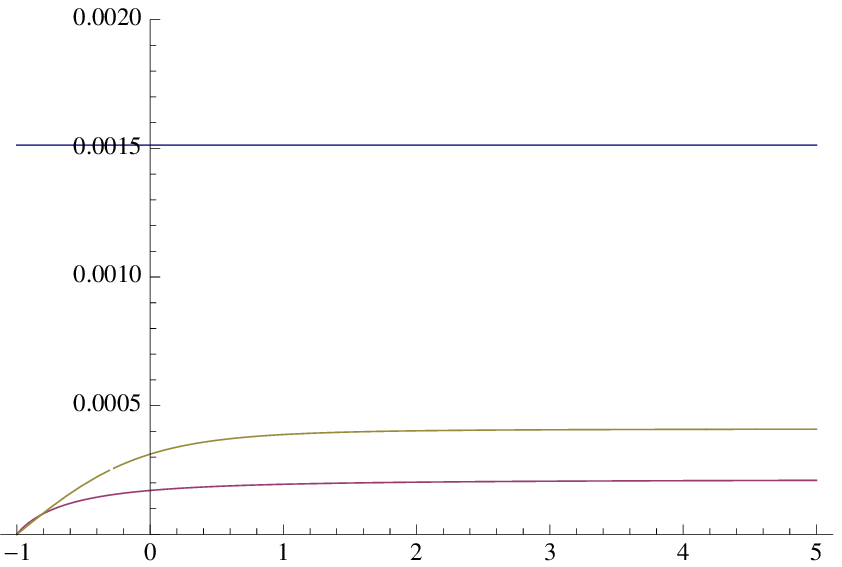,width=2.2truein,height=2.15truein}}
\caption{{\bf Left panel:} the density contrast as a function of redshift for the Einstein-de Sitter solution (blue), the spatially flat $\Lambda$CDM model with $\Omega_m = 0.27$ (yellow) and the present model with $\Omega_m = 0.45$ (red), for the same initial condition $\delta \approx 10^{-5}$ for $z = 1100$. {\bf Right panel:} the corresponding gravitational potential as a function of $z$, for $k=0.002\,h/$Mpc.}
\label{figdcontrast}
\end{figure*}

The only important consequence of matter production on the CMB spectrum could be owing to the integrated Sachs-Wolf effect (ISW).
In the right panel of Figure \ref{figdcontrast} we plot the gravitational potential $k^2\Phi = \rho_m a^2 \delta/2$ ($\delta$ is the matter density contrast and $k$ is the comoving wavenumber) as a function of the redshift, for the Einstein-de Sitter, $\Lambda$CDM and $\Lambda(t)$CDM models, for $k=0.002\,h/$Mpc. The ISW effect is  proportional to the integrated change of the gravitational potential. As the right panel of Fig.~\ref{figdcontrast} shows, the slope of the potential of our model differs from the corresponding slope of the $\Lambda$CDM model. But also here one may expect this effect to be small. In the $\Lambda$CDM model the potential decreases by $24$\% between the
last scattering and the present time, in our $\Lambda(t)$ model
it decreases by $20$\%. Note, however, that Fig.~\ref{figdcontrast} is preliminary since it is based on a Newtonian analysis \cite{Julio}. Here, a more quantitative, fully relativistic analysis is necessary.

On the other hand, as can also be seen in Fig.~\ref{figdcontrast}, there is a small difference in the gravitational potential as compared to the standard model, a difference that will affect the normalization of the CMB spectrum, since the Sachs-Wolf effect is proportional to $\Phi$. But the gravitational potential is proportional to the  matter density contrast as well. Therefore, if the latter is correctly normalized (by using, for example, the BBKS transfer function \cite{BBKS}, as was done, e.g., in \cite{Julio} and in the left panel of Fig.~\ref{figpowersp}), the CMB spectrum will also have the correct normalization. In other words, in order to have the correct normalization today, we have more power in the CMB spectrum at high redshifts as compared to the standard case.

\section{The coincidence problem}

The coincidence problem within the $\Lambda$CDM model is usually phrased as: why $\rho_{m}$ and $\Lambda$
are of the same order of magnitude just at the present epoch? Our model relies on dark-matter particle production at a constant rate. From the outset, this rate is a free parameter. By Eq.~(\ref{rate}) it is assumed here to be of the order of the Hubble constant. For our model the coincidence problem translates into: why the rate $\Gamma$ is of the order of the present-time Hubble rate $H_{0}$ (and not of the order of the Hubble rate at a different time)? Any other value of $\Gamma$ would not be compatible with the observed cosmological dynamics. 

Compared with the $\Lambda$CDM model, the coincidence problem is alleviated, however, in the sense that $\Omega_{m}/\Omega_{\Lambda} \propto a^{-3/2}$ \cite{Zimdahl}, whereas in the $\Lambda$CDM model one has $\Omega_{m}/\Omega_{\Lambda} \propto a^{-3}$. Any decay of the ratio $\Omega_{m}/\Omega_{\Lambda}$ that is less than $a^{-3}$ can be considered as an alleviation of the coincidence problem \cite{dalal}. 

This difference in the background dynamics also influences the behavior of the perturbations. As the left panel of Fig.~\ref{figdcontrast} shows, the present model deviates from an Einstein-de Sitter universe already at higher redshifts than the $\Lambda$CDM model. Moreover, while the perturbations within the latter approach a constant value for decreasing $z$, in our model there appears a maximum around the present epoch and perturbations are suppressed for $z \rightarrow -1$. 

\section{Conclusions}

The origin and nature of the dark sector is probably the central theoretical problem of modern cosmology. The non-zero tiny value of the cosmological term and its coincidence with the matter density are some important faces of the question. Relating dark energy with vacuum fluctuations is an old difficulty, but one may expect a potential solution if this problem is appropriately considered in curved spacetime. Several models of dynamic vacuum have been suggested along these lines. Models for  accelerated expansion based on matter creation in the expanding background have previously been proposed, e.g., in \cite{winfried,Ademir}. Although a relation between dynamical dark energy and particle production is not mandatory, we think it represents an interesting option.

This paper is based on the assumption that the late-time spacetime expansion can extract non-relativistic particles from the quantum vacuum. In the case of a constant creation rate, the backreaction of this process leads to a vacuum density linearly proportional to the Hubble parameter, a result previously obtained from estimations of the gravitational contribution of the QCD vacuum condensate. The resulting cosmological model is shown to be in accordance with the most precise observational tests, namely those which involve SNIa, the anisotropy spectrum of the CMB, the LSS distribution and BAO, albeit for higher values of the current matter density than in the standard model. Another difference to the $\Lambda$CDM model may be a modified ISW effect. This will be a subject of future research. 

\section*{Acknowledgements}

This work was partially supported by CNPq and FAPERJ.

{}

\end{document}